\documentclass[a4paper,12pt]{article}
\usepackage{amssymb}
\usepackage{amsmath}
\usepackage{amstext}
\usepackage{amsfonts}
\usepackage{epsfig}
\usepackage{dcolumn}
\usepackage{rotating}
\usepackage{color}
\usepackage{comment}
\usepackage{hyperref}

\def\XXint#1#2#3{{\setbox0=\hbox{$#1{#2#3}{\int}$ }
\vcenter{\hbox{$#2#3$ }}\kern-.56\wd0}}

\newcommand*\xbar[1]{%
  \hbox{%
    \vbox{%
      \hrule height 0.5pt % The actual bar
      \kern0.5ex%         % Distance between bar and symbol
      \hbox{%
        \kern-0.1em%      % Shortening on the left side
        \ensuremath{#1}%
        \kern-0.1em%      % Shortening on the right side
      }%
    }%
  }%
}

\definecolor{rosso}{cmyk}{0,1,1,0.4}
\definecolor{rossos}{cmyk}{0,1,1,0.55}
\definecolor{rossoc}{cmyk}{0,1,1,0.2}
\definecolor{blu}{cmyk}{1,1,0,0.3}
\definecolor{blus}{cmyk}{1,1,0,0.6}
\definecolor{bluc}{cmyk}{1,1,0,0.1}
\definecolor{verde}{cmyk}{0.92,0,0.59,0.25}
\definecolor{verdec}{cmyk}{0.92,0,0.59,0.15}
\definecolor{verdes}{cmyk}{0.92,0,0.59,0.7}

\newcommand{\ba}{\begin{eqnarray}}
\newcommand{\ea}{\end{eqnarray}}
\newcommand{\be}{\begin{equation}}
\newcommand{\ee}{\end{equation}}
\newcommand{\bi}{\begin{itemize}}
\newcommand{\ei}{\end{itemize}}
\newcommand{\al}{\alpha}
\newcommand{\bt}{\beta}
\newcommand{\ga}{\gamma}

\newcommand{\la}{\lambda}

\newcommand{\sa}{\sigma}
\newcommand{\en}{\epsilon}

\newcommand{\Ga}{\Gamma}

\newcommand{\La}{\Lambda}
\newcommand{\cF}{{\cal F}}

\newcommand{\w}{\widetilde}

\newcommand{\st}{\stackrel}

\newcommand{\ra}{\rightarrow}

\newcommand{\LF}{\left(}
\newcommand{\RF}{\right)}
\newcommand{\LT}{\left[}
\newcommand{\RT}{\right]}

\newcommand{\kb}{\bar{k}}
\newcommand{\pb}{\bar{p}}
\newcommand{\4}{\frac{1}{4}}

\newcommand{\mt}{\mathtt}

\newcommand{\non}{\nonumber\\}

\usepackage[normalem]{ulem}

\begin{document}

%+Title
\title{Towards UV Finiteness of Infinite Derivative Theories of Gravity and Field Theories}
\author{Spyridon Talaganis \\ \\
 {\it Consortium for Fundamental Physics,} \\
{\it Lancaster University, Lancaster,} \\
{\it LA$1$ $4$YB, United Kingdom.}\\
\begin{footnotesize}\textit{E-mail}:  s.talaganis@lancaster.ac.uk \end{footnotesize}}

\date{}

\maketitle
%-Title

%+Abstract
\begin{abstract}
In this paper we will consider the ultraviolet (UV) finiteness of the most general one-particle irreducible ($1$PI) Feynman diagrams within the context of ghost-free, infinite-derivative scalar toy model, which is inspired from ghost free and singularity-free infinite-derivative theory of gravity. We will show that by using dressed vertices and dressed propagators, $n$-loop, $N$-point diagrams constructed out of lower-loop $2$- \& $3$-point and, in general, $N_i$-point diagrams are UV finite with respect to internal and external loop momentum. Moreover, we will demonstrate that the external momentum dependences of the $n$-loop, $N$-point diagrams constructed out of lower-loop $2$- \& $3$-point and, in general, $N_i$-point diagrams decrease {\it exponentially}
as the loop-order increases and the external momentum divergences are eliminated at sufficiently high loop-order. 
\end{abstract}
%-Abstract

%+Contents
\tableofcontents
%-Contents

\section{Introduction}
\numberwithin{equation}{section}

Formulating a completely successful theory of quantum gravity~\cite{Veltman:1975vx,dewittQG,DeWitt:2007mi} remains a goal to be achieved in theoretical physics. Renormalizability plays a very important role in establishing a consistent theory of quantum gravity. In four-dimensions, pure gravity is ultraviolet (UV) finite at $1$-loop order~\cite{tHooft:1974toh}. That is, at $1$-loop order, one-loop counterterms vanish on mass-shell.
Now, at $2$-loop order, pure gravity has a UV divergence~\cite{tHooft:1974toh,Goroff:1985sz,vandeVen:1991gw}. Since infinitely many local counterterms would be required to eliminate the divergences, pure gravity is said to be a \textit{non-renormalizable} theory. By virtue of being non-renormalisable, pure gravity, as given by the Einstein-Hilbert action, is not a quantum theory of gravity, but, rather, an effective field theory, valid at scales much less than $M_{P}\sim 2.4\times 10^{18}$~GeV.

Among many attempts to quantise gravity, \textit{i.e.}, string theory~\cite{Polchinski:1998rr}, loop quantum gravity~\cite{Ashtekar,Nicolai:2005mc}, causal set~\cite{Henson:2006kf},  one can see that \textit{nonlocality} is present in many of these theories of quantum gravity; for example, strings and branes are nonlocal by nature~\cite{Eliezer:1989cr}. In string field theory~\cite{Witten:1985cc,Siegel:1988yz}, nonlocality also plays an important role ($p$-adic strings~\cite{Freund:1987kt}, zeta strings~\cite{Dragovich:2007wb} and strings quantized on a random lattice~\cite{Douglas:1989ve}). This merits the question whether nonlocality is a fundamental feature of spacetime. As a result, the investigation of nonlocality in nature seems to be a worthy pursuit. Moreover, many of the attempts to quantise gravity were afflicted by classical singularities, ghosts or divergences in the UV. For example, see Refs.~\cite{Stelle:1976gc,Stelle:1977ry}, a fourth-order higher-derivative theory of gravity was formulated that was renormalizable by power counting. However, the theory suffered from a massive spin-$2$ ghost, therefore lacking prediction both at a classical and at a quantum level.
In Ref.~\cite{VanNieuwenhuizen:1973fi}, a ghost-free tensor Lagrangian was presented and relevant applications in gravity were discussed. Nevertheless, any higher derivative extension of gravity would inevitably suffer from either ghost or classical instability due to the Ostrogradsky theorem~\cite{Label1}, which cannot be cured order by order in curvature corrections.

In fact, it has recently been shown how the ghost problem can be avoided in the context of \textit{infinite-derivative} theories of gravity~\cite{Biswas:2011ar,Biswas:2013kla,BG,Moffat-qg} (see Refs.~\cite{Efimov,Addazi} for discussion of unitarity in nonlocal theories).
Infinite derivatives indeed modify the action, and the dynamical propagating degrees of freedom. However, with a judicious choice, it is possible to make sure that gravity in any dimensions retains the transverse and traceless degrees of freedom, \textit{i.e.}, spin-$2$ and spin-$0$ components. Similar conclusion was reached by analysing the Hamiltonian density and considering primary/secondary constraints~\cite{Tal}.
Inspired by these attempts, one would like to formulate a \textit{ghost-free, nonlocal}, \textit{infinite-derivative} gravitational action and construct a renormalisable theory of gravity; see Ref.~\cite{Kuzmin:1989sp} for earlier work in that direction. Since the interactions in such class of theories are all derivatives in nature, the interactions due to infinite covariant derivatives give rise to nonlocal interactions~\footnote{The free theory has no nonlocality.}.

Within the context of \textit{infinite-derivative} field theories and gravity, one should start by explaining first what is meant by ultraviolet (UV)  finiteness.
When a Feynman diagram is finite in the UV, it means that the corresponding Feynman integral is convergent in the UV, \textit{i.e.}, at very high energies (or short distances). That is, there are no UV divergences, with respect to the \textit{internal loop momentum} variable $k^{\mu}$~\footnote{We work in four-dimensional spacetime, $\mu=0,1,2,3$, and we use the $(-,+,+,+)$ metric signature.}. As far as renormalisability is concerned, this implies that no counterterms are required to cancel possible UV divergences. One should keep in mind that, if all UV divergences with respect to internal loop momenta, which arise in one-particle irreducible ($1$PI) diagrams, can be removed by adding to the action finitely many counterterms, renormalisability is automatically ensured. 

Nevertheless, one could also study UV finiteness with respect to \textit{external momenta}. For instance, when computing scattering diagrams, if the scattering diagram exhibits no external momentum growth, it is convergent in the UV, that is, for large external momenta. In that case, the corresponding cross section of the scattering diagram is finite and does not blow up when the external momenta become very large.

A covariant gravitational theory free from ghosts and tachyons around constant curvature backgrounds were derived in Refs.~\cite{Biswas:2011ar,Biswas:2013kla}. The form of the action $S$ is given by
\begin{align}\label{aglar}
S & = S _{EH}+S_Q \,, \\
S_{EH} & = \frac{1}{2} \int d^{4} x \, \sqrt{-g} M_{P}^{2}  R \,, \\
S_{Q} &= \frac{1}{2} \int d^{4} x \, \sqrt{-g} \LF R \cF_{1}(\bar{\Box})R+R_{\mu \nu} \cF_{2}(\bar{\Box})R^{\mu \nu}+R_{\mu \nu \la \sa} \cF_{3}(\bar{\Box})R^{\mu \nu \la \sa} \RF \,,
\end{align}
where $\bar{\Box}\equiv \Box/M^2$ and $M$ is the mass scale at which the nonlocal modifications become important. The $\cF_{i}$'s are infinite-derivative functions of $\bar{\Box}$ and follow a specific constraint $2\cF_1(\bar{\Box})+\cF_2(\bar{\Box})+2\cF_3(\bar{\Box})=0$ around a Minkowski background so that the action is~\textit{ghost-free} and corresponds to a massless graviton~\cite{Biswas:2011ar,Biswas:2013kla}. In particular, in the action considered by Biswas, Gerwick, Koivisto and Mazumdar (BGKM), the graviton propagator is modulated by the exponential of an \textit{entire function} $a(-k^2) = e^{k^2/M^2}$, see~\cite{Biswas:2005qr}~\footnote{Note that a similar action has been proposed in Ref.~\cite{Tomboulis}, where it was shown that $a(-k^2) $ being an {\it entire} function is sufficient condition for the renormalisability of infinite derivative gravity. Later on similar conclusions were made in Ref.~\cite{Modesto}. },
\be \label{spyros}
\Pi(-k^2) = \frac{1}{k^2a(-k^2)}\LF {\cal P}^2 - \frac{1}{2} {\cal P}_s ^0  \RF=\frac{1}{a(-k^2)} \Pi_{GR} \,.
\ee
Note that the exponential of an entire function does not give rise to poles. For an exponential entire function, the propagator becomes exponentially suppressed in the UV (see also applications regarding Regge behaviour~\cite{Biswas:2004qu} and Hagedorn transition~\cite{Biswas:2009nx}), while the vertex factors are exponentially enhanced. The fact that the propagators and vertex factors have opposing momentum dependence is a key feature of gauge theories. Therefore, the UV divergences
of Feynman diagrams can be eliminated up to $2$-loop order~\cite{Talaganis:2014ida} and the theory is renormalisable. Higher loops can also be made finite by the use of dressed vertices and dressed propagators. Meanwhile, in the IR, we recover the physical graviton propagator of GR. In addition, this asymptotically free theory addresses the classical singularities present in GR~\cite{Biswas:2011ar,Biswas:2014tua}. This is in clear contrast with GR and other finite-order higher-derivative theories of gravity.

The gravitational entropy of the BGKM action for $D$-dimensional (A)dS backgrounds and the graviton propagator for the BGKM action around $D$-dimensional Minkowski space were evaluated in Refs.~\cite{Conroy:2015wfa,Conroy:2015nva}. In Ref.~\cite{Teimouri:2016ulk}, the generalised boundary term, \textit{i.e.}, the corresponding Gibbons-Hawking-York term, for the BGKM action was derived. Also, in Ref.~\cite{Tal}, the Hamiltonian for an infinite-derivative extension of gravity was presented and the number of degrees of freedom in various cases was computed.

Inspired by this infinite-derivative gravitational action, see Eq.~\eqref{aglar}, we have formulated a scalar toy model in Ref.~\cite{Talaganis:2014ida} that captures the essential features of the UV behaviour of the infinite-derivative gravitational action.
The scalar {\it toy model} action was given by 
\be \label{dacho}
S_{\mt{scalar}} = S _ {\mt{free}} + S _ {\mt{int}}\,,
\ee
where
\be
S_{\mt{free}} = \frac{1}{2}\int d^4 x \, \LF  \phi \Box a(\bar{\Box}) \phi\RF
\ee
and
\be
S_{\mt{int}} = \frac{1}{M_P} \int d ^ 4 x \, \LF \frac{1}{4} \phi \partial _ {\mu} \phi \partial ^ {\mu} \phi + \frac{1}{4} \phi \Box \phi a(\bar{\Box}) \phi - \frac{1}{4} \phi \partial _ {\mu} \phi a(\bar{\Box})  \partial ^ {\mu} \phi \RF\,;
\ee
we have $a (\bar{\Box}) = e^{- \bar{\Box}} \equiv e^{-\Box/M^2}$. The equation of motion for the action given by Eq.~\eqref{dacho} satisfies the shift-scaling symmetry $\phi \ra (1 + \epsilon) \phi + \epsilon$, where $\en$ is infinitesimal.

We should point out that in Ref.~\cite{Talaganis:2014ida} we showed that the infinite-derivative scalar toy model given by~\eqref{dacho} is renormalisable at $1$-loop order. That is, we computed the counterterm that cancels the UV divergences which originate from integrating the loop momentum variable $k^{\mu}$ at $1$-loop, \textit{i.e.}, for the $1$-loop, $2$-point function. In particular, in Ref.~\cite{Talaganis:2014ida}, we have shown that
\begin{itemize}
\begin{comment}
\item the BRST transformations and the invariance of our infinite-derivative gravitational action under those transformations,
\end{comment}
\item the Feynman rules for propagators and vertices for our infinite-derivative scalar toy model, that scattering amplitudes for the scalar toy model are superficially convergent for $L>1$, where $L$ is the number of loops, that the highest divergence of the $1$-loop, $2$-point function with nonvanishing external momenta $p$ \& $-p$ is $\La^4$, where $\La$ is a hard cutoff, and that the highest divergence of the $2$-loop, $2$-point function with vanishing external momenta is also $\La^4$,
\item the dressed propagator is more exponentially suppressed than the bare propagator and that dressed vertices behave as exponentials of external momenta when the external momenta are large; by employing dressed propagators and dressed vertices, $n$-loop, $2$- \& $3$-point diagrams constructed out of lower-loop, $2$- \& $3$-point diagrams become finite in the UV (UV) with respect to internal loop momentum $k^{\mu}$, that is, no UV divergences arise and no new counterterm is required.
\end{itemize}

In Ref.~\cite{Talaganis:2016ovm} the UV behaviour of scattering diagrams within the context of an infinite-derivative scalar toy model was investigated and it was found that the external momentum dependence of the scattering diagrams is convergent for large external momenta. That was achieved by dressing the bare vertices  of the scattering diagrams by considering renormalised propagator and vertex loop corrections to the bare vertices.
As the loop order increases, the exponents in the dressed vertices decrease and eventually become negative at sufficiently high loop-order.

Motivated by the results in Refs.~\cite{Talaganis:2014ida,Talaganis:2016ovm}, we would like to study UV finiteness with respect to both \textit{internal loop momenta} and \textit{external momenta} for a general class of Feynman diagrams within the context of infinite-derivative field theories. Thus, we shall generalise the results presented in Ref.~\cite{Talaganis:2014ida} and show that by employing dressed propagators and dressed vertices, $n$-loop, $N$-point diagrams constructed out of lower-loop $2$- \& $3$-point and, in general, $N_i$-point diagrams are finite in the UV (with respect to internal loop momentum $k^{\mu}$) while the exponential momentum dependences in those diagrams decrease at higher loops and external momentum divergences are eliminated at sufficiently high loop-order. It should be pointed out that $n$-loop, $N$-point diagrams constructed out of lower-loop $N_i$-point diagrams are the most general one-particle irreducible ($1$PI) Feynman diagrams.
In particular, we present the following results:
\begin{itemize}
\item the $1$-loop, $2$-point function with external momenta $p$ \& $-p$, where the bare propagators have been replaced with dressed propagators, is finite in the UV with respect to internal loop momentum, that is, the corresponding Feynman integrals are convergent,
\item by employing dressed vertices and dressed propagators, $n$-loop, $N$-point diagrams constructed out of lower-loop $2$- \& $3$-point and, in general, $N_i$-point diagrams are UV finite with respect to internal loop momentum,
\item the external momentum dependences of $n$-loop, $N$-point diagrams constructed out of lower-loop $2$- \& $3$-point and, in general, $N_i$-point diagrams decrease
as the loop-order increases and the external momentum divergences are eliminated at sufficiently high loop-order.
\end{itemize}

The outline of this paper is as follows. In section~\ref{sec:feynman}, we write down the Feynman rules, \textit{i.e.}, propagators and vertex factors, for the infinite-derivative scalar toy model. In section~\ref{sec:nmnm}, we show that the $1$-loop, $2$-point function with external momenta $p$ \& $-p$, where the bare propagators have been replaced with dressed propagators, is UV finite with respect to internal loop momentum $k^{\mu}$. In section~\ref{uvf}, we show the UV finiteness with respect to internal loop momentum of $n$-loop, $N$-point diagrams constructed out of lower-loop $2$- \& $3$-point and, in general, $N_i$-point diagrams. Moreover, we demonstrate that the external momentum dependences of $n$-loop, $N$-point diagrams constructed out of lower-loop $2$- \& $3$-point and, in general, $N_i$-point diagrams decrease
as the loop-order increases and, at sufficiently high loop-order, the external momentum divergences are eliminated. Finally, in section~\ref{sec:concl} we conclude by summarising our results. In appendix~\ref{sec:b}, we present some technical details regarding the $c_N$ coefficients.

\section{Feynman rules for infinite derivative scalar toy model}
\numberwithin{equation}{section}
\label{sec:feynman}
%%%%%%%%%%%%%%%%%%%%%%%%%%%

%%%%%%%%%%%%%%%%%%%%%%%%%%
All the Feynman rules and Feynman integral computations in this paper are carried out in Euclidean space after analytic continuation ($k_{0} \to i k _ {0}$ \& $k^2 \to k_{E}^2$ using the mostly plus metric signature; we shall drop the $E$ subscript for notational simplicity).

The Feynman rules for our action, which is given by Eq.~\eqref{dacho}, can be derived rather straightforwardly. The propagator in momentum space is then given by
\be
\Pi (k ^ 2)= \frac{- i}{k^2 e ^ {\kb ^ 2}}\,,
\ee
where barred $4$-momentum vectors from now on will denote the  momentum divided by the mass scale $M$. The vertex factor for three incoming momenta $k_{1},~k_{2},~k_{3}$ satisfying the conservation law:
\be
k _ {1} + k _ {2} + k _ {3} = 0\,,
\label{conservation}
\ee
is given by
\be
\label{eq:V}
\frac{1}{M_{P}}V (k _ {1}, k _ {2}, k _ {3}) = \frac{i}{M_P} C(k_1,k_2,k_3) \LT 1 -  e ^ {\kb _ {1} ^ {2}} -  e ^ {\kb _ {2} ^ {2}} - e ^ {\kb _ {3} ^ {2}}\RT\,,
\ee
where
\be
C (k_1,k_2,k_3)= \frac{1}{4} \LF k _ {1} ^ {2} + k _ {2} ^ {2} + k _ {3} ^ {2} \RF\,.
\ee
Let us briefly explain how we obtain the vertex factor.
The first term originates from the term, $ \4 \phi \partial _ {\mu} \phi \partial ^ {\mu} \phi$, which using Eq.~\eqref{conservation} in the momentum space, reads
\ba
- \frac{i}{2} (k _ {1} \cdot k _ {2} + k _ {2} \cdot k _ {3} + k _ {3} \cdot k _ {1})
=  \frac{i}{4} \left(k _ {1} ^ {2} + k _ {2} ^ {2} + k _ {3} ^ {2} \right) \,.
\ea
The second term comes from the terms, $\frac{1}{4} \phi \Box \phi a(\Box) \phi$, and $- \frac{1}{4} \phi \partial _ {\mu} \phi a(\Box)  \partial ^ {\mu} \phi$. In the
momentum space, again using Eq.~\eqref{conservation}, we get
\be
\frac{i}{4} \left(k _ {3} \cdot k _ {1} + k _ {1} \cdot k _ {2} - k _ {3} ^ {2} - k _ {2} ^ {2} \right) e ^ {\kb _ {1} ^ {2}}  = - \frac{i}{4} \left(k _ {1} ^ {2} + k _ {2} ^ {2} + k _ {3} ^ {2} \right) e ^ {\kb _ {1} ^ {2}} \,.
\ee
 The third and the fourth terms in Eq.~\eqref{eq:V} arise in an identical fashion.

\section{$1$-loop, $2$-point function with non-vanishing external momenta}
\numberwithin{equation}{section}
\label{sec:nmnm}

\begin{figure}[t]
\centering
\includegraphics[width=1.0\textwidth]{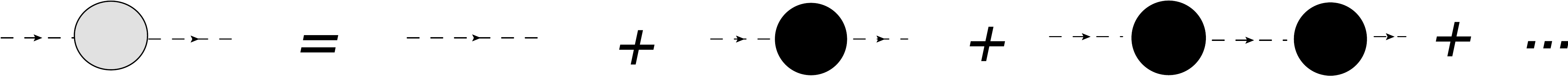}
\caption{\label{fig:dressed} {\small The dressed propagator as the sum of an infinite geometric series. The dressed propagator is denoted by the shaded blob.}}
\end{figure}

\begin{figure}[t]
\centering
\includegraphics[width=0.3\textwidth]{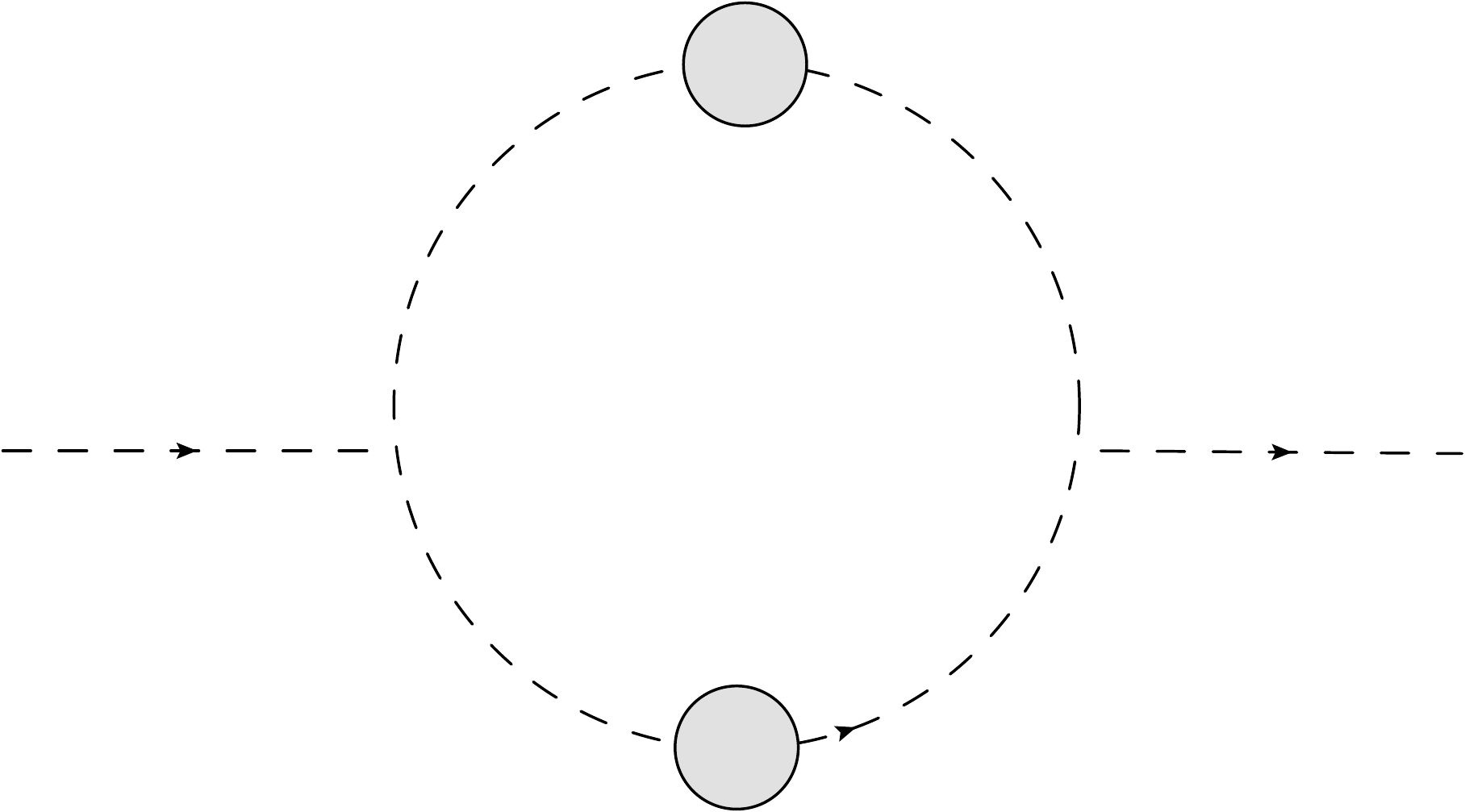}
\caption{\label{fig:odra} {\small The $1$-loop, $2$-point function with dressed propagators. The shaded blobs denote dressed propagators.}}
\end{figure}

From~\cite{Talaganis:2014ida} we know that the renormalised $1$-loop, $2$-point function with external momenta $p$ and $-p$, $\Ga_{2,1\mt{r}}(p^2)$, is a regular analytic function of $p^2$ which grows as $e^{3\pb^2/2}$ as $p^2 \ra \infty$. Within the context of dimensional regularization, the counterterm $\Ga_{2,1\mt{ct}}$ has a simple pole in $\en$, where $\en=4-d$ and $d$ is the dimensionality of spacetime.  That is,
\be
\Ga_{2,1\mt{r}}(p^2)=\Ga_{2,1}(p^2)+\Ga_{2,1\mt{ct}}(p^2)=\frac{i M^4}{M_P^2}f(\pb^2)
\ee
and, as $p^2 \ra \infty$, 
\be
f(\pb^2)\sim e^{\frac{3\pb^2}{2}} \,.
\ee

The dressed propagator then represents the geometric series of all the graphs with $1$-loop, $2$-point insertions as shown in Fig.~\ref{fig:dressed}, analytically continued to the entire complex $p^2$-plane.  Mathematically, this is equivalent to replacing the bare propagator, $\Pi(p^2)$, with the dressed propagator, $\w{\Pi}(p^2)$:
\be \label{gako}
\w{\Pi}(p^2)= \frac{\Pi(p^2)}{1-\Pi(p^2)\Ga_{2,1\mt{r}}(p^2)}= \frac{1}{i \LF p ^ {2 } e ^ { \bar{p} ^ {2}}-\frac{M^4}{M_P^2}f\LF\bar{p}^2\RF \RF}\,.
\ee
Since in our case $\Pi(p^2)\Ga_{2,1\mt{r}}(p^2)$ grows with large momenta, in the UV limit we have
\be
\label{dressed-UV}
\w{\Pi}(p^2)\ra \Ga^{-1}_{2,1\mt{r}}(p^2)\approx \LF 9-12 \pb^{-2}  \RF ^ {-1} e^{-\frac{3\bar{p}^2}{2}}\,.
\ee
We observe that the dressed propagator is more exponentially suppressed than the bare propagator. 

Now if we replace the bare propagators with dressed propagators in the $1$-loop, $2$-point function with external momenta $p$ \& $-p$ while the vertices stay bare, the Feynman integral, see Fig.~\ref{fig:odra}, is given by
\begin{align}\label{plpo}
\Ga_{2,1\mt{dressed}}(p^2)& = \frac{1}{2 i  M _ {p} ^ {2}} \int \frac{d ^ {4} k}{(2 \pi) ^ {4}} \, \frac{V ^ {2} (-p, \frac{p}{2} + k, \frac{p}{2} - k)}{\LT (\frac{p}{2}+k ) ^ {2}e ^ {\left(\frac{\pb}{2}+\kb \right) ^ {2}}-\frac{M^4}{M_P^2}f \LF ( \frac{\pb}{2}+\kb )^{2}\RF \RT} \non
& \times \frac{1}{\LT (\frac{p}{2}-k ) ^ {2}e ^ {\left(\frac{\pb}{2}-\kb \right) ^ {2}}-\frac{M^4}{M_P^2}f \LF ( \frac{\pb}{2}-\kb )^{2}\RF \RT} \,.
\end{align}
As $\lvert k \rvert \ra \infty$, the integrand goes as
\be
\sim \exp (-k^2) \,.
\ee
Therefore, the integral is convergent since there are no internal loop momentum divergences. On the other hand, we observe that $\Ga_{2,1\mt{dressed}}$ goes, in terms of external momentum $p$, as \be
\sim e^{\frac{5\pb^2}{4}}
\ee 
for large $p^2$. We observe that the exponential momentum dependence of $\Ga_{2,1\mt{dressed}}$
is less divergent than the exponential momentum dependence of the renormalised $1$-loop, $2$-point function $\Ga_{2,1\mt{r}}(p^2)$.
We would like to make our theory not only renormalisable (apart from the $1$-loop, $2$-point function with bare propagators, all other higher-loop, higher-point diagrams can be made UV finite with respect to internal loop momentum), but also UV finite with respect to external momenta (at sufficiently high loop-order).

Therefore, in the next section we shall dress both vertices and propagators so as to make $n$-loop, $N$-point diagrams UV finite, both with respect to internal loop momentum and external momenta.

\section{UV finiteness of $n$-loop, $N$-point diagrams}
\numberwithin{equation}{section}
\label{uvf}

We would like to investigate the UV finiteness of one-particle irreducible ($1$PI) Feynman diagrams within the framework of our infinite-derivative scalar toy model. We shall look into UV finiteness both in terms of internal loop momenta and external momenta. In the next section,  we shall look into the UV finiteness of $n$-loop, $N$-point diagrams constructed out of lower-loop $2$- \& $3$-point and, in general, $N_i$-point diagrams; it should be pointed out that $n$-loop, $N$-point diagrams constructed out of lower-loop $N_i$-point diagrams are the most general one-particle irreducible ($1$PI) diagrams.

\begin{figure}[t]
\centering
\includegraphics[width=.40\textwidth]{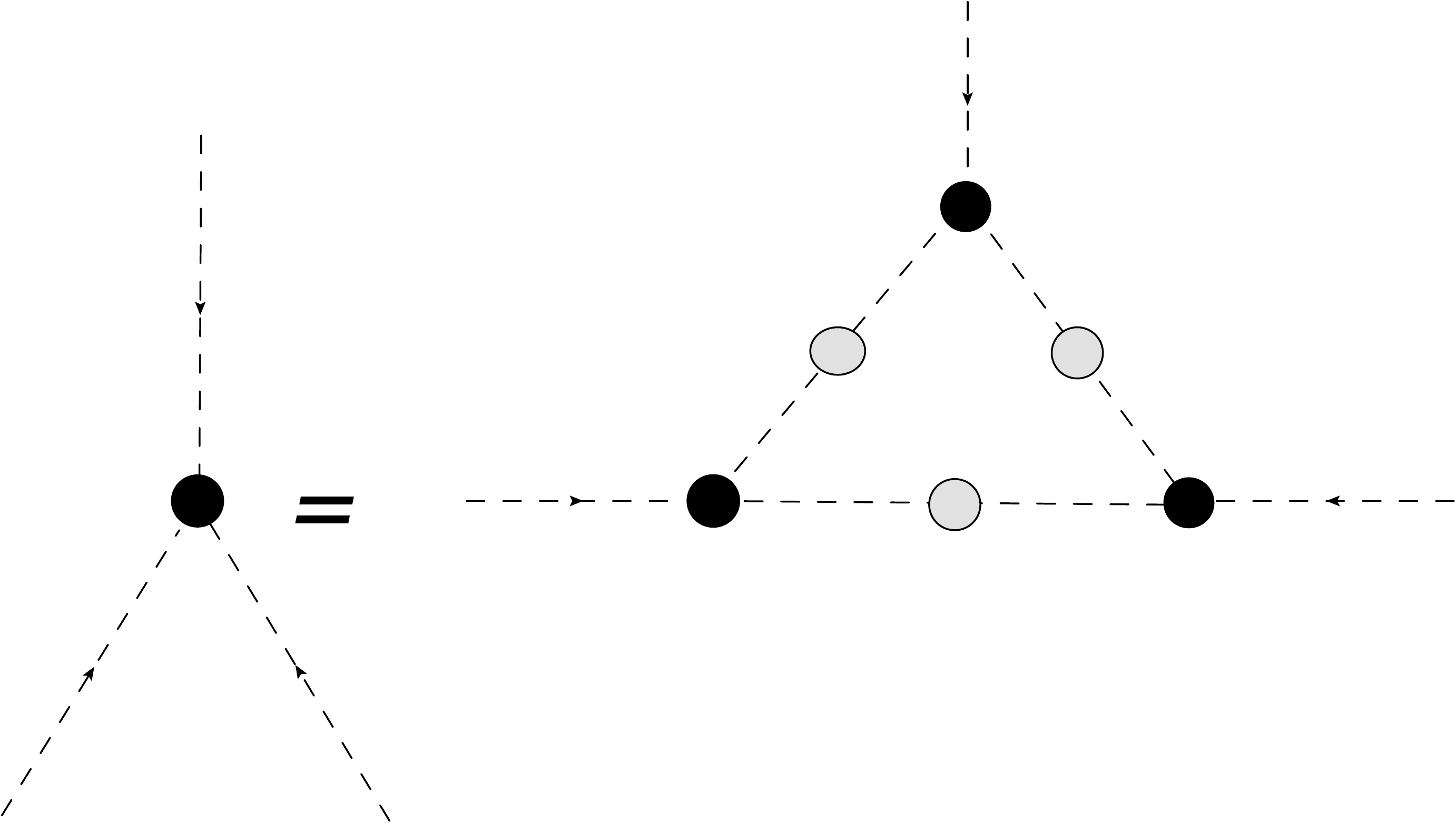}
\caption{\label{set} {\small $3$-point diagram constructed out of lower-loop $2$-point \& $3$-point diagrams. The shaded blobs indicate dressed internal propagators and the dark blobs indicate renormalised vertex corrections. The loop order of the dark blob on the left is $n$ while the loop order of the dark blobs on the right is $n-1$. The external momenta are $p_1$, $p_2$, $p_3$  and the internal (that is, inside the loop) momenta are $k+\frac{p_{1}}{3}-\frac{p_2}{3}$, $k+\frac{p_{2}}{3}-\frac{p_3}{3}$, $k+\frac{p_{3}}{3}-\frac{p_1}{3}$.}}
\end{figure}

\begin{figure}[t]
\centering
%\includegraphics[width=.40\textwidth]{N-Polygon_dressed.pdf}
%\hspace{1cm}
\includegraphics[width=.40\textwidth]{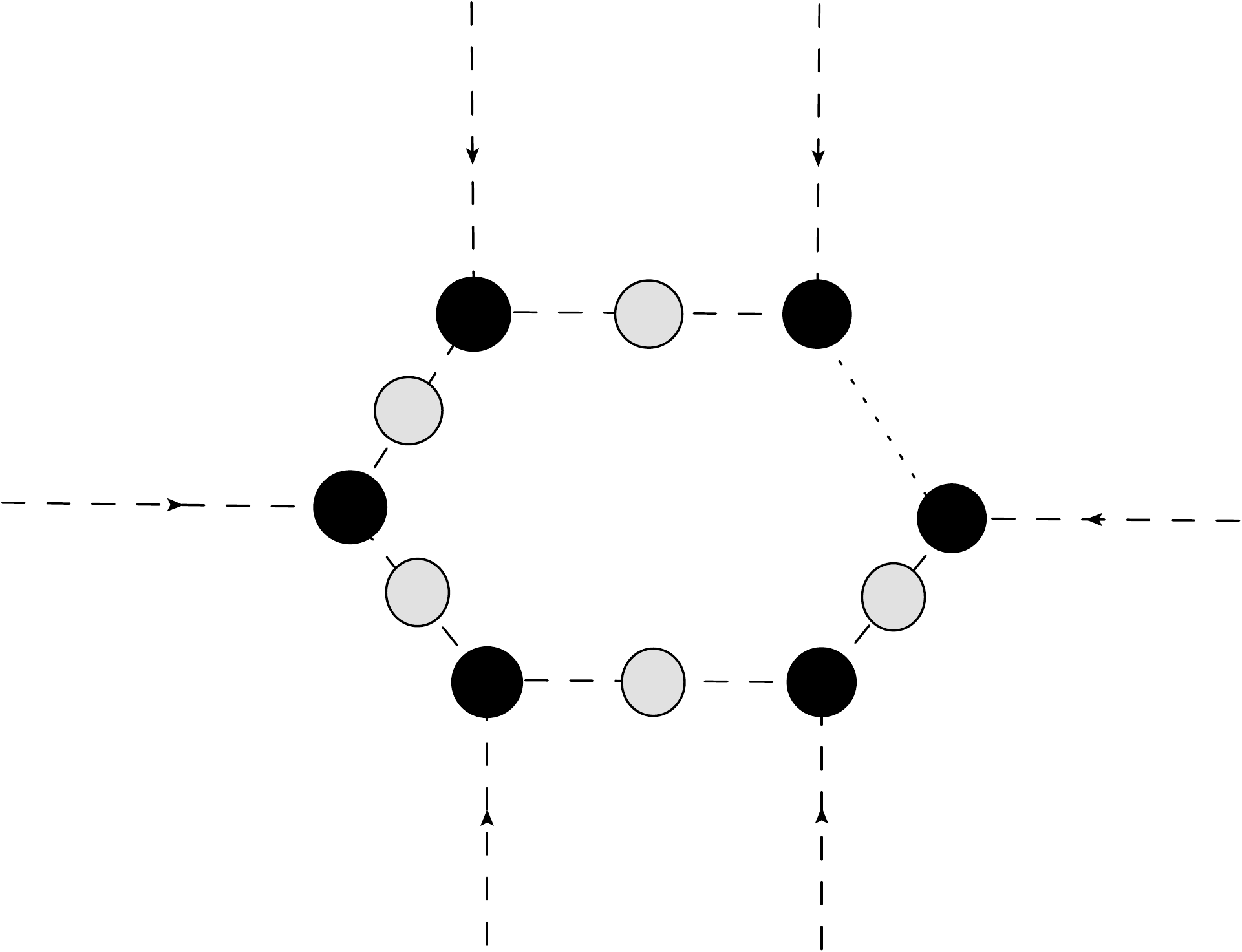}
\caption{\label{fig:Ndressed} {\small $n$-loop, $N$-point diagram constructed out of lower-loop $2$- \& $3$-point diagrams with loop corrections to the vertices (dark blobs) and dressed propagators (shaded blobs). The internal dots indicate an arbitrary number of renormalised vertex corrections and dressed propagators.}}
\end{figure}

\subsection{$n$-loop, $N$-point diagrams constructed out of lower-loop, $2$- \& $3$-point diagrams}
\label{sec:oda}

 Let us look at $N$-point diagrams constructed out of lower-loop $2$- \& $3$-point diagrams. We know the following: 
\begin{itemize}
\item the dressed propagators represented by the shaded blobs decay in the UV as $e^{-{3\kb^2}/{2}}$ (see Eq.~\eqref{dressed-UV} in section~\ref{sec:nmnm}).
\item the $3$-point function represented by the dark blobs (see Figs.~\ref{set}~\&~\ref{fig:Ndressed}) can be written (after first integrating out the internal loop momentum $k^{\mu}$ in the $1$-loop triangle) as
\be
\label{eq:lola}
\Ga_3\st{UV}{\longrightarrow}\sum_{\al,\bt,\ga} e^{\al \pb_{1}^{2}+\bt \pb_{2}^{2}+\ga \pb_{3}^{2}} \,,
\ee
with  the convention (since we assume $p_{1}+p_{2}+p_{3}=0$, terms such as $p_{i}\cdot p_{j}$, where $i,j=1,2,3$, can be written as a sum of $p_l^2$, $l=1,2,3$, terms)
\be
\al\geq\bt\geq \ga \,,
\ee
where $p_1,p_2 ,p_3$ are the three external momenta. 
\end{itemize}

Now one can generalise Eq.~\eqref{eq:lola} and write the $N$-point function in the following form (again after first integrating out the internal loop momentum $k^{\mu}$ in the $1$-loop $N$-polygon, see also appendix~\ref{sec:b}):
\be
\label{eq:dola}
\Ga_N\st{UV}{\longrightarrow}\sum_{\al_l} e^{\sum_{l=1}^{N}\al_{l}\pb_{l}^{2}} \,,
\ee
with  the convention
\be
\al_{1}\geq\al_{2}\geq \dots \geq \al_{N} \,,
\ee
where $p_1,p_2,\dots,p_N$ are the $N$ external momenta.

Now let us look at the case where the external momenta are arbitrary; we wish to find out how one can get the largest exponents for the external momenta. First, let us consider how one can get the largest sum of all the exponents, {\it i.e.}, $\sum_{l=1}^{N}\al_{l}$. Even though all the arguments below can be conducted for $N$ different sets of exponents in the $N$ $3$-point vertices, see Fig.~\ref{set}, making up the $1$-loop $N$-polygon, see Fig.~\ref{fig:Ndressed}, for simplicity, here we will look at what happens when all the $N$ vertices have the same exponents. The best way to obtain the largest exponents for the external momenta is to have the $\al$ exponent correspond to the external momenta (we assume a symmetric distribution of $(\bt, \ga)$ among the internal loops and symmetrical routing of momenta in the $1$-loop $N$-polygon).
We have that $p_{1},~p_{2},\dots,~p_{N}$ are the external momenta for the $1$-loop triangle, and  the superscript in the $\al^{n-1},\bt^{n-1},\ga^{n-1}$ indicates that these are coefficients that one obtains from contributions up to $n-1$  loop level. The internal momenta of the $N$-point diagram are given by
\begin{align}
q_{N-1} &= \frac{1}{N}\LT \sum_{l=1}^{N-2}\LF lp_{l} \RF -p_{N-1} \RT \,, \non
q_{N} &= \frac{1}{N}\LT \sum_{l=1}^{N-2}\LF lp_{l+1} \RF -p_{N} \RT \,, \non
& \vdots \non
q_{N-3} &= \frac{1}{N}\LT p_{N-1}+2p_{N}+ \sum_{l=1}^{N-4}\LF (l+2)p_{l} \RF -p_{N-3} \RT \,, \non
q_{N-2} &= \frac{1}{N}\LT p_{N}+ \sum_{l=1}^{N-3}\LF (l+1)p_{l} \RF -p_{N-2} \RT \,.
\end{align}
That is, 
\begin{itemize}
\item the dressed propagators are given by $e^{-\frac{3}{2}(\kb+\bar{q}_{l})^{2}}$, $l=1,\dots,N$, 
\item and the vertex factors are of the form $e ^ {\al^{n-1} \pb _ {l} ^ {2} + \bt^{n-1} \LF \kb+\bar{q}_{l}  \RF^2 + \ga^{n-1} \LF \kb+\bar{q}_{l+
1} \RF ^2}$. 
\end{itemize}
Hence, conservation of momenta then yields
\be
\label{eq:crucial98}
\Ga_{N,n}{\longrightarrow}\int \frac{d ^ 4 k}{(2 \pi) ^ 4} \frac{e^{\al^{n-1}(\pb_1^2+\pb_2^2+\dots+\pb_N^2)}}{e^{[\frac{3}{2}-\bt^{n-1}-\ga^{n-1}][N\kb^2+c_{N}(\pb_1^2+\pb_2^2+\dots+\pb_N^2)]} }\,,
\ee
where $c_{N}$ is a coefficient depending on $N$, the number of external lines, that satisfies $c_{N}>0$ for all $N$ (see appendix~\ref{sec:b}).

\begin{itemize}
\item \textit{Internal momentum}: we observe that the integrand in~\eqref{eq:crucial98} is of the form $e^{-s\kb^2}$, where $s>0$ (in Ref~\cite{Talaganis:2014ida}, it was shown that $\bt^{n-1}+\ga^{n-1}<\frac{3}{2}$ for all $n$). Hence, the integral in~\eqref{eq:crucial98} is convergent and the diagram is UV finite with respect to internal loop momentum.
\item \textit{External momenta}: by integrating Eq. \eqref{eq:crucial98}, we have
\be \label{boko}
\al_1^n=\al_2^n=\dots=\al_N^n=\al^{n-1}+c_{N}(\bt^{n-1}+\ga^{n-1})-\frac{3c_{N}}{2} \,.
\ee
In particular, for the $1$-loop, $3$-point graph, one has to use the $3$-point bare vertices: $\al^0=1$ and $\bt^0=\ga^0=0$. In Ref.~\cite{Talaganis:2016ovm} we showed that the coefficients $\al^{n-1}$, $\bt^{n-1}$, $\ga^{n-1}$, that is, the exponents in the dressed vertices, decrease as the loop-order increases and, at sufficiently high loop-order, become negative. In particular, we showed in Ref.~\cite{Talaganis:2016ovm} that, for $n=1$ ($n$ is the loop-order of the $3$-point dressed vertices),
\be \label{rulo}
\al^1=\bt^1=\ga^1=\frac{1}{2}\,,
\ee
for $n=2$,
\be
\al^2=\bt^2=\ga^2=\frac{1}{3}\,,
\ee
for $n=3$,
\be
\al^3=\bt^3=\ga^3=\frac{1}{18}\,,
\ee
for $n=4$,
\be
\al^4=\bt^4=\ga^4=-\frac{11}{27}\,.
\ee
We conclude that, for $n \geq 4$, $\al^n$, $\bt^n$ and $\ga^n$ become negative.

Hence, from Eq.~\eqref{boko} we see that the coefficients $\al_1^n,~\al_2^n,\dots,\al_N^n$ also decrease as the loop-order increases and, at sufficiently high loop-order, become negative. Thus, the external momentum dependences of $n$-loop, $N$-point diagrams constructed out of lower-loop $2$- \& $3$-point diagrams decrease
as the loop-order increases and the external momentum divergences are eliminated at sufficiently high loop-order, that is, the exponents in Eq.~\eqref{eq:dola} corresponding to the external momenta become negative when the loop order is sufficiently large.
\end{itemize}

 Consequently, this class of diagrams are finite in the UV, both with respect to internal loop momentum and at sufficiently high loop-order external momenta as well. One could also consider the case where the loop-order of the dressed vertices is not the same for all of the dressed vertices, that is, each dressed vertex is of different loop-order. Again the results would be the same as far as UV finiteness with respect to internal loop momentum and external momenta is concerned.

As a check, when the external momenta tend to zero, it is easy to see that the most divergent UV part of the $N$-point diagram reads
\be
\label{eq:514}
\Ga_{N,n}{\longrightarrow}\int \frac{d ^ 4 k}{(2 \pi) ^ 4} \, \frac{e^{(\al_1+\dots+\al_N+\bt_1+\dots+\bt_N)\kb^{2}}}{e^{\frac{3N\kb^{2}}{2}}}\,,
\ee
where $k$ is the loop momentum variable in Fig.~\ref{fig:Ndressed}. There are $N$ propagators $e^{\frac{3 \kb^2}{2}}$   while the (most divergent UV parts of the) vertex factors originating from lower-loop diagrams are $e ^ {\al_{1} \kb^2 + \bt_{1}\kb^2}$,\dots,$e ^ {\al_{N} \kb^2 + \bt_{N}\kb^2}$ (we get no $\ga_{1}$, $\ga_{2}$ terms in the exponents, since the external momenta are set equal to zero). Clearly, the integral is finite as long as
\be
\al_{i}+\bt_{i}<\frac{3}{2}\,,
\label{condition1}
\ee
where $i=1, 2,\dots,N$. 

\subsection{$n$-loop, $N$-point diagrams constructed out of lower-loop, $N_i$-point diagrams}
\label{sec:koda}

\begin{figure}[t]
\centering
%\includegraphics[width=.40\textwidth]{N-Polygon_dressed.pdf}
%\hspace{1cm}
\includegraphics[width=.40\textwidth]{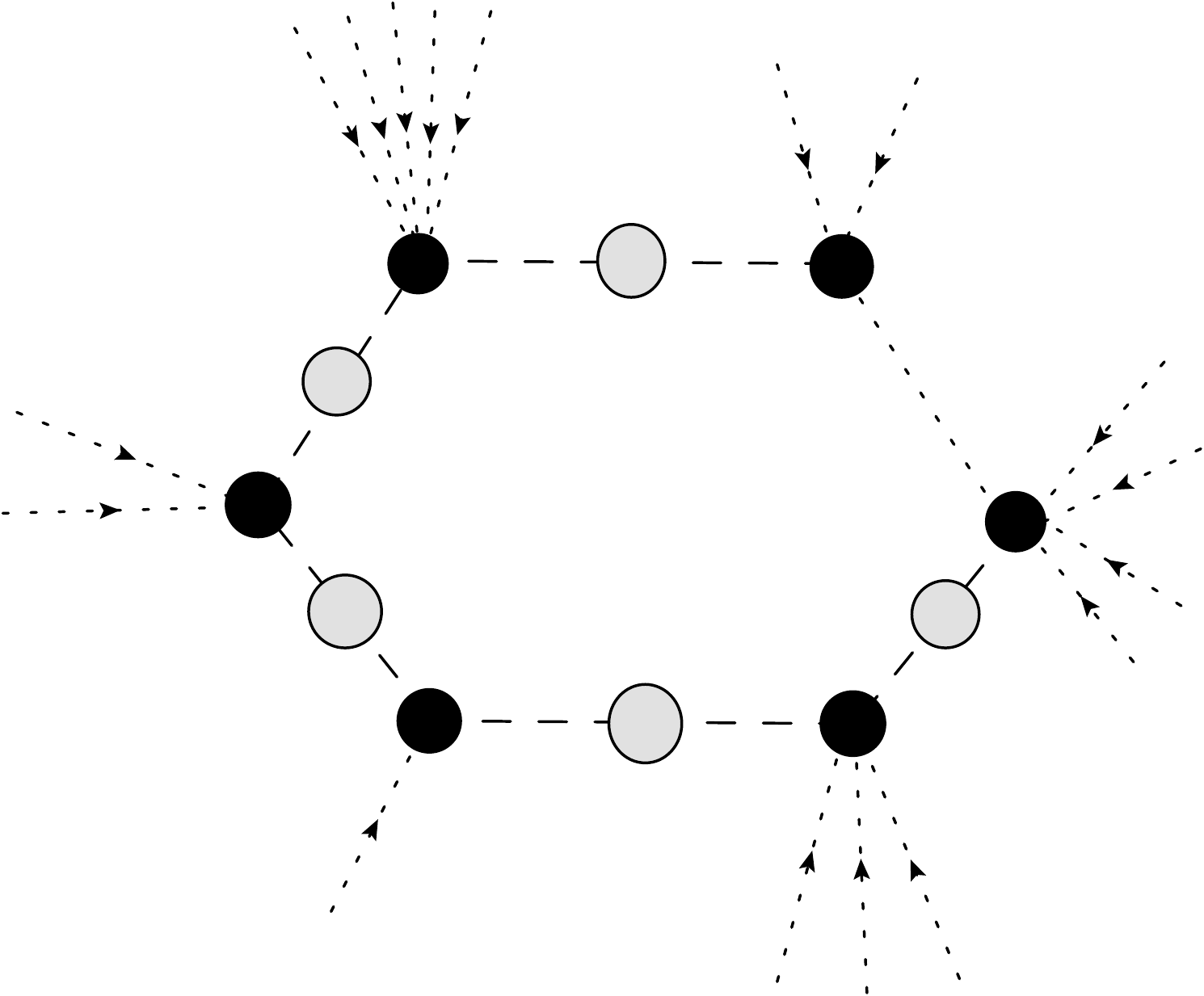}
\caption{\label{fig:Ndressedo} {\small $n$-loop, $N$-point diagram constructed out of lower-loop $N_i$-point diagrams with loop corrections to the vertices (dark blobs) and dressed propagators (shaded blobs). The internal dots indicate an arbitrary number of renormalised vertex corrections and dressed propagators. The external dots indicate an arbitrary number of external lines.   }}
\end{figure}

Now let us look at $n$-loop, $N$-point diagrams constructed out of lower-loop $N_{i}$-point diagrams. Any $n$-loop diagram can be thought of as a $1$-loop integral over a graph containing renormalised vertex corrections and dressed propagators, see Fig.~\ref{fig:Ndressedo}. At the $i$-th dressed vertex, we have $N_{i}-2$ external lines (excluding the two internal propagators in the $n$-loop, $N$-point diagram that are attached to each dressed vertex).

For simplicity, we take all the vertices to have the same exponents. We have $m$ dressed vertices with $N_i$ external lines attached to each dressed vertex; thus,
\be
N=\sum_{i=1}^{m}(N_{i}-2) \,.
\ee 
Regarding the external momenta, we use the following notation:
\be
p_{i}^{'}=\sum_{l=1}^{N_i-2} p_{i_l} \,,
\ee
where $i=1,\dots,m$ and $p_{i_l}$ are the external momenta to the $i$-th dressed vertex. For each dressed vertex, the $N_i$-point function can be written as follows:
\be
\label{doko}
\Ga_{N_{i}}\st{UV}{\longrightarrow}\sum_{\al_{i_l},\bt,\ga} e^{\sum_{l=1}^{N_i-2} \al_{i_l}\pb_{i_l}^{2}+\bt_{i} \bar{q}_{1}^{2}+\ga_{i} \bar{q}_{2}^{2}}\,,
\ee
where $q_1$ and $q_2$ are internal propagators in the $N$-point diagram, with the convention
\be
\al_{i_1}\geq \al_{i_2}\geq \dots \geq \al_{i_{N_{i}-2}} \geq \bt_{i} \geq \ga_{i} \,.
\ee

The internal momenta of the $N$-point diagram are given by
\begin{align}
q_{m-1}^{'} &= \frac{1}{m}\LT \sum_{l=1}^{m-2}\LF lp_{l}^{'} \RF -p_{m-1}^{'} \RT \,, \non
q_{m}^{'} &= \frac{1}{m}\LT \sum_{l=1}^{m-2}\LF lp_{l+1}^{'} \RF -p_{m}^{'} \RT \,, \non
& \vdots \non
q_{m-3}^{'} &= \frac{1}{m}\LT p_{m-1}^{'}+2p_{m}^{'}+ \sum_{l=1}^{m-4}\LF (l+2)p_{l}^{'} \RF -p_{m-3}^{'} \RT \,, \non
q_{m-2}^{'} &= \frac{1}{m}\LT p_{m}^{'}+ \sum_{l=1}^{m-3}\LF (l+1)p_{l}^{'} \RF -p_{m-2}^{'} \RT \,.
\end{align}
That is, 
\begin{itemize}
\item the dressed propagators are given by $e^{-\frac{3}{2}(\kb+\bar{q}_{i}^{'})^{2}}$, $i=1,\dots,m$, 
\item and the vertex factors are of the form $e^{\sum_{l=1}^{N_i-2} \al_{i_l}^{n-1}\pb_{i_l}^{2}+ \bt_{i}^{n-1} \LF \kb+\bar{q}_{l}^{'}  \RF^2 + \ga_{i}^{n-1} \LF \kb+\bar{q}_{l+
1}^{'} \RF ^2}$. 
\end{itemize}

Hence, conservation of momenta then yields (after shifting the loop momentum variable $k$)
\be
\label{eq:crucial99}
\Ga_{N,n}{\longrightarrow}\int \frac{d ^ 4 k}{(2 \pi) ^ 4} \frac{e^{\sum_{i=1}^{m}\sum_{l=1}^{N_i-2} \al_{i_l}^{n-1}\pb_{i_l}^{2}}}{e^{[\frac{3m}{2}-\sum_{i=1}^{m}(\bt_{i}^{n-1}+\ga_{i}^{n-1})]\kb^{2}} e^{\sum_{i=1}^{m}\sum_{l=1}^{N_i-2} [\frac{3}{2}b_{i_l}-\bt_{i}^{n-1}c_{i_l}-\ga_{i}^{n-1}d_{i_l}]\pb_{i_l}^{2}}}\,,
\ee
where $b_{i_l}$,~$c_{i_l}$~\&~$d_{i_l}$ are coefficients depending on $i$ \& $l$ that satisfy $b_{i_l},c_{i_l},d_{i_l}>0$ for all values of $i$ and $l$. 
\begin{itemize}
\item \textit{Internal momentum}: we observe that the integrand in~\eqref{eq:crucial99} is of the form $e^{-s\kb^2}$, where $s>0$ (that is, $\sum_{i=1}^{m}(\bt_{i}^{n-1}+\ga_{i}^{n-1})<\frac{3m}{2}$). Hence, the integral in~\eqref{eq:crucial99} is convergent and the diagram is UV finite with respect to internal loop momentum.
\item \textit{External momenta}: by integrating Eq. \eqref{eq:crucial98}, we have
\be \label{bloko}
\al_1^n=\al_2^n=\dots=\al_N^n=\al_{i_l}^{n-1}+c_{i_l}\bt_{i}^{n-1}+d_{i_l}\ga_{i}^{n-1}-\frac{3b_{i_l}}{2} \,.
\ee
From section~\ref{sec:oda}, one can see that the coefficients $\al_{i_l}^{n-1}$, $\bt_{i}^{n-1}$, $\ga_{i}^{n-1}$, that is, the exponents in the dressed vertices, decrease as the loop-order increases and at sufficiently high loop-order become negative. 
Hence, from Eq.~\eqref{bloko} we see that the coefficients $\al_1^n,~\al_2^n,\dots,\al_N^n$ also decrease as the loop-order increases and at sufficiently high loop-order become negative. Thus, the external momentum dependences of $n$-loop, $N$-point diagrams constructed out of lower-loop $N_i$-point diagrams decrease
as the loop-order increases and the external momentum divergences are eliminated at sufficiently high loop-order, that is, the exponents in Eq.~\eqref{doko} corresponding to the external momenta become negative when the loop order is sufficiently large.
\end{itemize}

Now, if each dressed vertex were of different loop-order, we would still obtain the same results regarding UV finiteness with respect to internal loop momentum and external momenta. Hence, $n$-loop, $N$-point diagrams constructed out of lower-loop $N_i$-point diagrams are finite in the UV, both with respect to internal loop momentum and at sufficiently high loop-order external momenta as well.

\begin{comment}

\subsection{$n$-loop, $N$-point diagrams constructed out of lower-loop, $N$-point diagrams}

Let us look at $n$-loop, $N$-point diagrams constructed out of lower-loop, $N}$-point diagrams. We shall demonstrate the finiteness of higher loops using a recursive argument.

\end{comment}

%We want to check whether the gauge-fixing and ghost-antighost terms contribute %to the propagator. Moreover, we wish to ascertain whether the ghosts \& %antighosts affect the superficial degree of divergence and the evaluation %of loop integrals in the UV.

\section{Conclusions}
\numberwithin{equation}{section}
\label{sec:concl}

The aim of this paper has been to show that $n$-loop, $N$-point diagrams constructed out of lower-loop $2$- \& $3$-point and, in general, $N_i$-point  diagrams are UV finite with respect to internal loop momentum, that is, when computing the Feynman integrals for those diagrams, no UV divergences arise and no new counterterm is required.
At the beginning we presented our infinite-derivative scalar toy model, which was inspired from an infinite-derivative theory of gravity, BGKM gravity, and wrote down the Feynman rules, that is, the propagator and the vertex factors. Next we summarised the results for the $1$-loop, $2$-point function with non-vanishing external momenta; when we replaced the bare propagators with dressed propagators in the $1$-loop, $2$-point function, we saw that the corresponding Feynman integrals were convergent. 

Then we showed that by employing dressed vertices and dressed propagators, $n$-loop, $N$-point diagrams constructed out of lower-loop $2$- \& $3$-point and, in general, $N_i$-point diagrams were UV finite. This is the major result in our paper: that the most general one-particle irreducible ($1$PI) Feynman diagrams within the framework of infinite-derivative field theories are finite in the UV. Consequently, no UV divergences arise and no new counterterm is required.

Furthermore, we showed that the external momentum dependences of $n$-loop, $N$-point diagrams constructed out of lower-loop $2$- \& $3$-point and, in general, $N_i$-point diagrams decrease
as the loop-order increases and the external momentum divergences are eliminated at sufficiently high loop-order.

Finally we would like to to further generalise the results obtained in this paper and apply them to an infinite-derivative theory of gravity, \textit{i.e.}, to BGKM gravity. Establishing that an infinite-derivative theory of gravity is renormalisable or even finite at higher loop-order would be a big achievement by itself and a major contribution to establishing a theory of quantum gravity that is successful on all fronts.

\section{Acknowledgments}

ST is supported by a scholarship from the Onassis Foundation. 

\setcounter{equation}{0}
\appendix

\section{$c_N$ coefficients}
\label{sec:b}

The $c_N$ coefficients are always positive. For instance, when $N=4$, \textit{i.e.}, for an $n$-loop, four-point diagram constructed out of lower-loop, three-point diagrams, we have, from the internal propagators and dressed vertices comprising the $n$-loop, four-point diagram, that (assuming symmetrical routing of momenta in the $1$-loop square)
\begin{align}\label{duro}
&\LF \kb+\frac{\pb_1}{4}+\frac{2\pb_2}{4}-\frac{\pb_3}{4} \RF^{2}+\LF \kb+\frac{\pb_2}{4}+\frac{2\pb_3}{4}-\frac{\pb_4}{4} \RF^{2}+\LF \kb+\frac{\pb_3}{4}+\frac{2\pb_4}{4}-\frac{\pb_1}{4} \RF^{2}\non
+&\LF \kb+\frac{\pb_4}{4}+\frac{2\pb_1}{4}-\frac{\pb_2}{4} \RF^{2}\non
=&4\kb^{2}+\frac{3}{8}\LF \pb_{1}^{2}+\pb_{2}^{2}+\pb_{3}^{2}+\pb_{4}^{2} \RF-\frac{1}{4}\LF p_{1}\cdot p_{3}+p_{2}\cdot p_{4} \RF \,.
\end{align} 
Now even if $p_{1}=p_{3}=p$ and $p_{2}=p_{4}=-p$, Eq.~\eqref{duro} is equal to 
\be
4\kb^{2}+\frac{1}{4}\LF \pb_{1}^{2}+\pb_{2}^{2}+\pb_{3}^{2}+\pb_{4}^{2} \RF \,.
\ee
We see that the coefficient $\frac{1}{4}$ is greater than zero. Now, when $N=5$, \textit{i.e.}, for an $n$-loop, five-point diagram constructed out of lower-loop three-point diagrams, we have (again assuming symmetrical routing of momenta in the $1$-loop pentagon)
\begin{align}\label{kuro}
&\LF \kb+\frac{\pb_1}{5}+\frac{2\pb_2}{5}+\frac{3\pb_3}{5}-\frac{\pb_4}{5} \RF^{2}++\LF \kb+\frac{\pb_2}{5}+\frac{2\pb_3}{5}+\frac{3\pb_4}{5}-\frac{\pb_5}{5} \RF^{2}+\LF \kb+\frac{\pb_3}{5}+\frac{2\pb_4}{5}+\frac{3\pb_5}{5}-\frac{\pb_1}{5} \RF^{2}\non
+&\LF \kb+\frac{\pb_4}{5}+\frac{2\pb_5}{5}+\frac{3\pb_1}{5}-\frac{\pb_2}{5} \RF^{2}+\LF \kb+\frac{\pb_5}{5}+\frac{2\pb_1}{5}+\frac{3\pb_2}{5}-\frac{\pb_3}{5} \RF^{2} \non
=& 5\kb^{2}+\frac{3}{5}\LF \pb_{1}^{2}+\pb_{2}^{2}+\pb_{3}^{2}+\pb_{4}^{2} \RF+\frac{2}{5}\LF p_{1}\cdot p_{2}+p_{2}\cdot p_{3}+p_{3}\cdot p_{4}+p_{4}\cdot p_{5}+p_{5}\cdot p_{1} \RF \,.
\end{align} 
Even when $p_{1}=2p$, $p_{2}=-p$, $p_{3}=p$, $p_{4}=-p$, $p_{5}=-p$, Eq.~\eqref{kuro} is equal to
\be
5\kb^{2}+\frac{7}{20}\LF \pb_{1}^{2}+\pb_{2}^{2}+\pb_{3}^{2}+\pb_{4}^{2}+\pb_{5}^{2} \RF \,.
\ee
Again the coefficient $\frac{7}{20}$ is greater than zero. One can proceed in a similar fashion for the other higher-point diagrams.

\end{document}